\documentclass[english,preprint2]{aa}
\usepackage{times}
\usepackage[T1]{fontenc}
\usepackage[latin1]{inputenc}
\usepackage{babel}
\usepackage{graphics}

\makeatletter

\providecommand{\LyX}{L\kern-.1667em\lower.25em\hbox{Y}\kern-.125emX\@}
\let\SF@@footnote\footnote
\def\footnote{\ifx\@typeset@protect
    \expandafter\SF@@footnote
  \else
    \expandafter\SF@gobble@opt
  \fi
}
\expandafter\def\csname SF@gobble@opt \endcsname{\@ifnextchar[
  \SF@gobble@twobracket
  \@gobble
}
\edef\SF@gobble@opt{\noexpand
  \expandafter\noexpand\csname SF@gobble@opt \endcsname}
\def\SF@gobble@twobracket[#1]#2{}

\usepackage{natbib}
\usepackage{indentfirst}

\makeatother
\begin{document}

\title{Additional TWA Members?}

\subtitle{Spectroscopic verification of kinematically selected TWA candidates}

\author{Inseok Song\inst{1}, M. S. Bessell\inst{2}, B. Zuckerman\inst{1}}

\institute{Department of Physics and Astronomy (UCLA), Los Angeles, CA 90095--1562,
USA\and Research School of Astronomy and Astrophysics, The Australian
National University, ACT 2611, Australia}

\offprints{Inseok Song}

\mail{song@astro.ucla.edu}

\date{Received: September, 2001 / Accepted: January, 2002} 
\abstract{We present spectroscopic measurements of the 23 new candidate
members of the TW~Hydrae Association from \citet{MF}. Based on H$\alpha $
and Li~6708~\AA{} strengths together with location on a color-magnitude
diagram for Hipparcos TWA candidates, we found only three possible new 
members (TYC~7760-0835-1, TYC~8238-1462-1, and TYC~8234-2856-1)
in addition to the already known member, TWA~19. This eliminated
most of the candidates more distant than 100~pc. Three Tycho stars,
almost certainly members of the Lower Centaurus Crux association, are the
most distant members of the TWA.  A claim of isotropic expansion
of TWA has to be re-evaluated based on our new results. Generally, one 
cannot identify new members of a diffuse nearby stellar group based solely 
on kinematic data. To eliminate
interlopers with similar kinematics, spectroscopic  verification is essential.
\keywords{open clusters and associations: individual: TWA -- Stars:
kinematics -- planetary systems -- Techniques: spectroscopic}}

\maketitle

\section{Introduction}

During the past few years, a handful of young ($<100$~Myrs)
and nearby ($\leq 100$~pc) groups of stars have been identified.
Usually, they occupy a few hundred square degrees in the plane of the sky
(thus are less conspicuous than compact clusters) and are
not associated with any known interstellar clouds. Therefore they have
been hard to find and have remained unnoticed until recently.
Beginning with the TW Hydrae Association \citep["TWA", ][and references therein]{TWA},
currently, there are about eight such associations reported in the
literature. For the most up-to-date information on these groups of
stars, readers are refered to \citet{YStars}. 

Thermal emission from massive planets can be detectable when they
are separated far enough from the primary stars. For a given linear separation
between primary and planet, a closer system to Earth will provide a larger 
angular separation making detection of the planet easier.
Planets around young stars are still hot and
they emit a fair amount of infrared radiation. Therefore, young and
nearby stars are excellent targets for observational searches for giant
planets and dusty ``proto-planetary'' disks.  The TWA (d$\sim 60$~pc),
the $\beta $~Pictoris moving group (d$\sim 35$~pc), and the Tucana/HorA
Association (d$\sim 45$~pc) are the three
youngest and closest groups to Earth; hence they form the best targets
for ground- and space-based searches for planets. The Capricornus group
\citep{vandenAncker} is also a very young nearby group, however
it is almost certainly a subgroup of the bigger $\beta $~Pictoris
group \citep{bPic2}. 

There are currently 19
known star systems in TWA and most of its members have been observed
with the  Hubble Space Telescope and/or with a ground-based telescope with
adaptive optics (AO) capability. Some TWA members possess substellar
companions (TWA~5; \citealt{TWA5}) or prominent dusty disks
(TW~Hydrae, HR~4796A, Hen3-600, and HD~98800). Any newly identified members 
of TWA, therefore, are excellent targets for planet or planetary disk searches.

\def\mmc{\multicolumn{2}{c}}
\begin{table*}

\caption{Spectroscopic data for Makarov and Fabricious TWA candidates.}

{\centering \begin{tabular}{clcr@{$\pm$}lr@{$\pm$}lr@{$\pm$}lcccc}
\hline 

Name&\multicolumn{1}{c}{ $(B-V)^{\dag}$}& R.V.& \multicolumn{6}{c}{R.V. measured (km/sec)}& EW(H$\alpha $)& EW(Li)& $v\sin i$& TWA\\
\cline{4-9} 
HIP or TYC&\multicolumn{1}{c}{(error)}& predicted&
\mmc{SSO}&
\mmc{Torres et al.}&
\mmc{Others$^{\ddag }$}& (\AA{})& (m\AA{})& (km/sec)& member?\\
\hline

53911      & 0.64(12)    & 12.7& \mmc{--}    & $+$12.9&0.3 & \mmc{--}    & $-220$  & 390  & 4  & TWA 1  \\
7201-0027-1& 1.88(37)    & 10.6& \mmc{--}    & $+$11.2&0.3 & \mmc{--}    & $-1.89$ & 490  & 13 & TWA 2  \\
55505      & 1.19(03)    &  9.1& \mmc{--}    &  $+$9.2&0.2 & \mmc{--}    & $0$     & 360  & -- & TWA 4  \\
7223-0275-1& 1.44(33)    & 10.0& $-$30.6&6.6 & $+$6.9:&2.0 & \mmc{--}    & $-13.4$ & 570  & 36:& TWA 5  \\
7183-1477-1& 1.61(26)    & 16.2& \mmc{--}    & \mmc{--}    & \mmc{--}    & $-4.65$ & 560  & -- & TWA 6  \\
7190-2111-1& 1.31(18)    & 11.0& \mmc{--}    & \mmc{--}    & \mmc{--}    & $-4.95$ & 440  & -- & TWA 7  \\
57589      & 1.66(40)$^G$& 12.6& \mmc{--}    & $+$10.2&0.4 & \mmc{--}    & $-5.01$ & 480  & -- & TWA 9  \\
61498      & 0.00(01)$^G$& 10.5& \mmc{--}    & \mmc{--}    &  $+$9.4&2.3 & --      & --   & -- & TWA 11 \\
57524      & 0.60(02)    & 20.0& $+$11.5&3.8 & \mmc{--}    & \mmc{--}    & $0.57$  & 189  & 24 & TWA 19 \\
\hline                   
46535      & 0.50(01)$^G$& 16.0& $+$18.0&7.0 & \mmc{--}    & $+$22.1&5.0 & $1.43$  & 63   & 37 & no     \\
47039      & 0.42(01)$^G$& 16.9& $+$10.9&1.3 & \mmc{--}    & $+$12.2&0.4 & $1.63$  & $<10$&  8 & no     \\
48273      & 0.48(01)$^G$& 10.7& $-$31.4&2.9 & $+$16.2&0.1 & \mmc{$+17.0^{\S }$}&2.0& 21 \& 22 & 9 \& 14& no\\
0829-0845-1& 0.69(05)    & 20.4&  $+$9.5&6.5 & \mmc{--}    & \mmc{--}    & $0.85$  & $<10$& 30 & no     \\
6604-0118-1& 1.07(06)    & 16.5&\mmc{$-19,+54,+69$}& $+$27.0&0.3 & \mmc{--}& $-0.5$& 93   & 20 & no     \\
49530      & 0.94(01)$^G$& 20.7& $+$22.5&1.1 & \mmc{--}    & $+$16.7&2.0 & $1.17$  & 34   & 2  & no     \\
6625-1087-1& 0.87(13)    & 21.3& $+$16.1&1.5 & \mmc{--}    & \mmc{--}    & $0.28$  & 141  & 10 & no     \\
7178-1493-1& 0.73(18)    & 15.2& $+$48.0&3.0 & \mmc{--}    & \mmc{--}    & $0.17$  & $<10$& 15 & no     \\
7183-1879-1& 0.96(20)    & 18.9& $+$11.0&1.6 & \mmc{--}    & \mmc{--}    & $0.74$  & 79   & 11 & no     \\
7188-0575-1& 1.11(05)    & 12.2& $+$42.9&2.1 & \mmc{--}    & \mmc{--}    & $-0.28$ & 50   & 25 & no     \\
50796      & 1.19(02)$^G$& 13.1& $+$22.4&0.9 & $+$13.1&1.0 & \mmc{--}    & $0.20$  & $<10$& 8  & no     \\
7710-2231-1& 1.07(03)    & 22.5& $-$12.2&1.0 & \mmc{--}    & \mmc{--}    & $0.63$  & $<14$& 7  & no     \\
52462      & 0.87(01)$^G$&  9.0& $+$22.7&0.5 & \mmc{--}    & \mmc{--}    & $0.78$  & 150  & 1  & no     \\
52787      & 0.82(02)    &  9.9& $+$24.0&0.6 & \mmc{--}    & \mmc{--}    & $0.82$  & $120$& 2  & no     \\
53486      & 0.91(02)$^G$&  3.7& $+$4.3 &1.0 &  $+$5.5&0.3 & \mmc{--}    & $0.88$  & $<10$& 1  & no     \\
55899      & 0.07(02)    & 21.8& \mmc{--}    & \mmc{--}    & \mmc{--}    & $2.0$   & --   & 210& no?    \\
57129      & 0.54(03)    & 12.9& $+$15&20    & \mmc{--}    & $+$25.6&10  & $1.56$  & $<10$& 190& no     \\
57269      & 0.91(01)$^G$&  8.5& \mmc{--}    & \mmc{--}    & \mmc{$+15.9^{\star }$}&--& 196& 20& no     \\
59315      & 0.71(01)$^G$&  6.6& $+$15.9&0.6 & \mmc{--}    & \mmc{--}    & $0.83$  & 151  & 5  & no     \\
7760-0835-1& 0.51(03)    & 19.4& $+$10.0&2.6 & \mmc{--}    & \mmc{--}    & $0.83$  & 161  & 12 & yes?   \\
8238-1462-1& 0.77(04)    & 16.8& $+$12.0&3.0 & \mmc{--}    & \mmc{--}    & $0.36$  & 294  & 18 & yes?   \\
8234-2856-1& 0.81(05)    & 20.1& $+$13.2&2.4 & \mmc{--}    & \mmc{--}    & $-0.49$ & 342  & 16 & yes?   \\
\hline
\multicolumn{13}{l}{$^{\dag }$ estimated from Tycho $(B_{T}-V_{T})$
using a relation given in \citet{Bessell}.}  \\
\multicolumn{13}{l}{~~ Stars with superscript "$G$", $(B-V)$ colors are from ground
photometry adapted from Hipparcos Input Catalog.}\\
\multicolumn{13}{l}{$^{\ddag }$ radial velocities from \citet{B-B} unless noted otherwise.}\\
\multicolumn{13}{l}{$^{\S }$ radial velocity from \citet{HIP48273}.}\\
\multicolumn{13}{l}{$\star $ radial velocity from \citet{HIP57269}.}\\
\end{tabular}\par}
\end{table*}

Using a modified convergent point method, \citet[MF hereafter]{MF}
found 23 new kinematic candidate members of TWA beside eight
known members. In fact, one of the 23 candidates is a known TWA member,
TWA~19 \citep[HIP\,57524;][]{TWA}. Based on a kinematic model of TWA including
the candidate members, MF predicted radial velocities of the 23 new
candidates. They also suggest that TWA expands with a rate of $0.12\, 
\mathrm{km}/\mathrm{sec}\cdot \mathrm{pc}^{-1}$
and that TWA is part of a larger structure, the ``Gould disk'', rather
than a separate group of stars. Considering the importance of TWA
members as aforementioned observational targets and as corner-stones
in investigations of the surrounding environment (local
association, Sco--Cen complex, etc.), all of the MF candidates need
to be verified spectroscopically. In this paper,
we present spectroscopic data for all MF candidates and discern true
members from interlopers based on spectroscopic youth indicators.

\section{Observations}

We are in the midst of an extensive survey to search for young and nearby
stars to Earth at Siding Spring Observatory (Australia) and Lick Observatory
(USA). We took spectra of all 23 MF candidates during January, April
and June 2001 and January 2002 observations at SSO. We used an 
echelle spectrograph on the
Nasmyth-B focus of the Australian National University's 2.3~m telescope.
Eight orders of the echelle spectrum cover portions of the wavelength range
from 5800~\AA{} to 7230~\AA{}. We focused on orders which contained
the H$\alpha $ and Li~6708~\AA{} lines. In these orders, the
measured resolution was 0.40~\AA{}. Radial velocities were determined
by cross-correlating target and radial velocity standard spectra over
5 or 6 orders of the echelle which were chosen to produce strong correlations
and have few atmospheric features. Projected rotational velocities
were measured from a total of about 10 lines in the Li~6708~\AA{} echelle
order with a procedure similar to that of \citet{Strassmeier}. Equivalent
widths of H$\alpha $ and Li~6708~\AA{} were measured using
the IRAF task \texttt{splot}. 

\citet{Torres} measured radial velocities of 9 stars, four (TWA~5,
TYC~6604-0118-1, HIP 48273 and HIP~50796) of which we also observed.
Three of the common stars (TWA~5, TYC~6604-0118-1, and HIP~48273) are known
binaries for which \citet{Torres} extracted binary orbits. The binarity
can explain discrepancies (Table~1) between our measured radial velocities and
those of \citet{Torres}. Our measured radial velocity of HIP~50796
differs from that of \citet{Torres} by $\sim 10$~km/sec. On
the night of our HIP~50796 observation, the extracted radial velocities
of two radial velocity standard stars were consistent with those from 
published catalogs within the errors ($\sim2$~km/sec). Thus,
we believe that our measurement is correct and HIP~50796 is likely
a spectroscopic binary also. Later HIP~50796 turned out to be a non-TWA
star. A list of 31 stars from MF with our spectroscopic data are
summarized in Table~1.

\section{Results. }

Because of the high fraction of binaries among young stars \citep[see][for example]{Patience98},
oftentimes, stellar radial velocities are affected by binary orbital
motions. Therefore, we cannot simply compare MF's predicted radial
velocities with measured ones to accept or reject the candidates.
In addition, MF assumed an isotropic expansion of TWA. If the assumption
is not correct, comparing predicted and measured radial velocities
is meaningless. 

{\bfseries
The very young age of TWA ($\leq 10$~Myrs) enables us to select
true TWA members based on their strong Li~6708~\AA{} absorption line 
strength depending on the spectral type of the candidates. By comparing
equivalent widths of Li~6708~\AA{} line for all candidates with those of 
comparably young open clusters members (Figure~1), we can eliminate more than 
half of the candidates. The eliminated candidates have too weak a Li~6708~\AA{} 
absorption line to be a $\leq10$~Myr old star. 

An H$\alpha$ feature (either in emission or partially filled-in) can be 
used to discern young stars from old stars. Generally, if a late--type star
that is not an interacting binary shows H$\alpha $ in emission or
partially filled-in, the star is young. However, even among the very young 
stars like the members of TWA, early type stars (earlier than mid-K) usually do
not show H$\alpha $ in emission. Unfortunately, most of the MF
candidates are earlier than mid-K, thus, we have not used H$\alpha $
as a tool to reject candidates. }

\begin{figure}
{\centering \resizebox*{0.9\columnwidth}{!}{\includegraphics{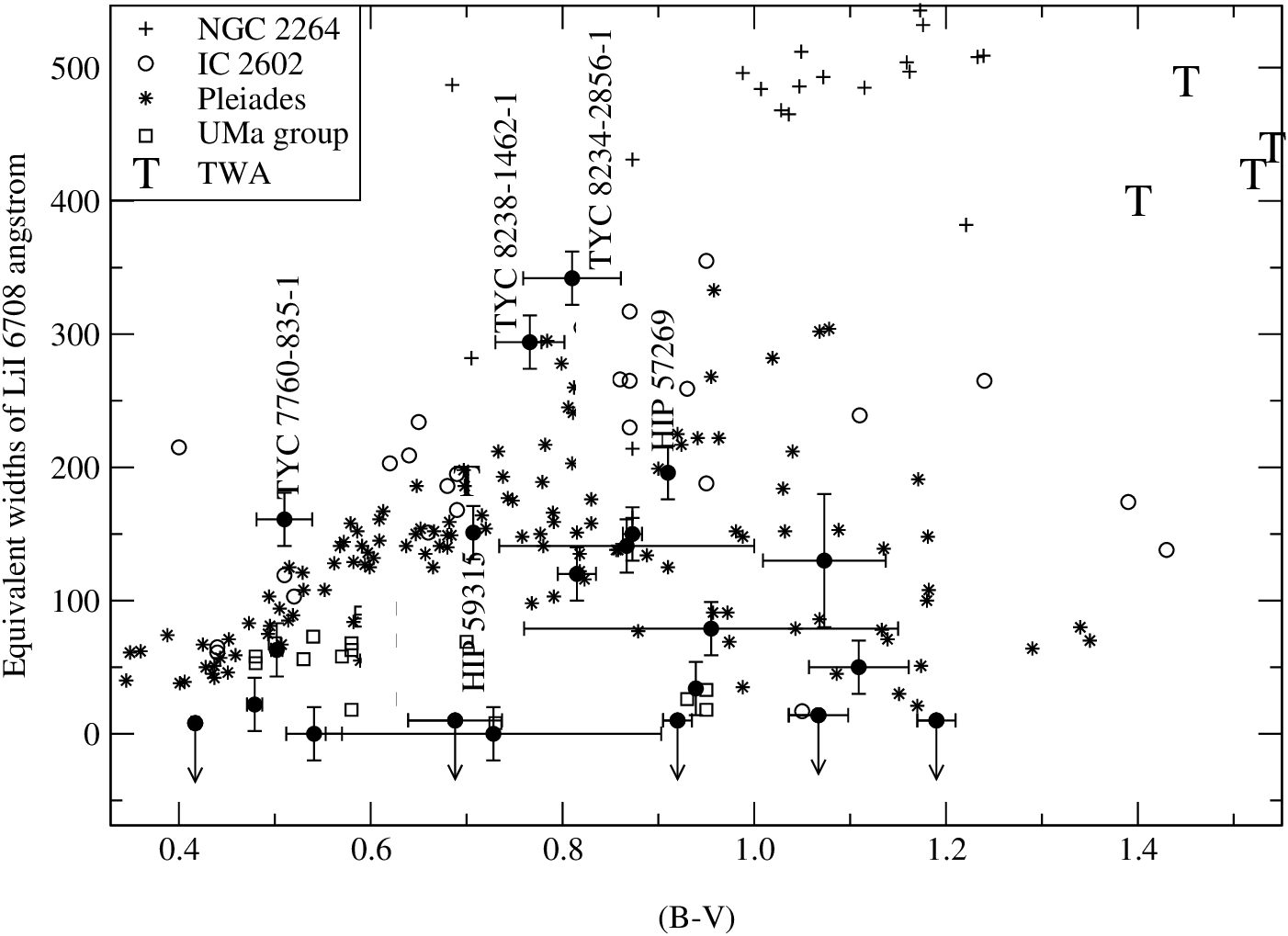}} \par}

\caption{Equivalent widths of Li~6708~\AA{} as a function of $(B-V)$.
Displayed equivalent widths are not corrected for the possible contamination
by FeI~6707.44~\AA{} and measurement uncertainty of EW(Li) is about
20~m\AA{}. Plus signs indicate NGC~2264 stars ($<10$~Myrs),
open circles denote IC~2602 stars ($10-30$~Myrs),
star symbols represent Pleiades stars ($\sim 125$~Myrs),
and open squares are Ursa~Majoris moving group stars ($\sim 300$~Myrs).
Previously known TWA members are plotted as the symbol `T' and solid circles
indicate MF candidates.}
\end{figure}

The age of TWA ($\sim 10$~Myrs) falls between those of NGC~2264
($<10$~Myrs) and IC~2602 ($10\sim 30$~Myrs). Equivalent
widths of previously known TWA members (symbol `T' in Figure~1) nicely
fall between the regions encompassing NGC~2264 and IC~2602. Therefore, 
we can use the lower envelope of IC~2602 Li~6708~\AA{} equivalent widths 
to reject false TWA members. Based on the Li~6708~\AA{}
equivalent widths, we eliminated TYC~0829-0845-1, TYC~6604-0118-1,
HIP~49530, TYC~6625-1087-1, TYC~7178-1493-1, TYC~7183-1879-1,
TYC~7188-0575-1, {\bfseries HIP~47039, HIP~48273}, HIP~50796, 
TYC~7710-2231-1, HIP~52462, HIP~52787, {\bfseries HIP~53486},
HIP~57129, HIP~57269, and HIP~59315. HIP~59315 and HIP~57269
have fairly large Li~6708~\AA{} equivalent widths and they may be
as young as $\sim 30$~Myrs, but their Li~6708~\AA{} equivalent
widths are too small to be included as TWA members. They are
also rejected on additional grounds (see below).

Independent age estimations were obtained for all Hipparcos MF candidates 
based on their location in a color--magnitude diagram (Figure~2).
\begin{figure}
{\centering \begin{tabular}{c}
\resizebox*{0.9\columnwidth}{!}{\includegraphics{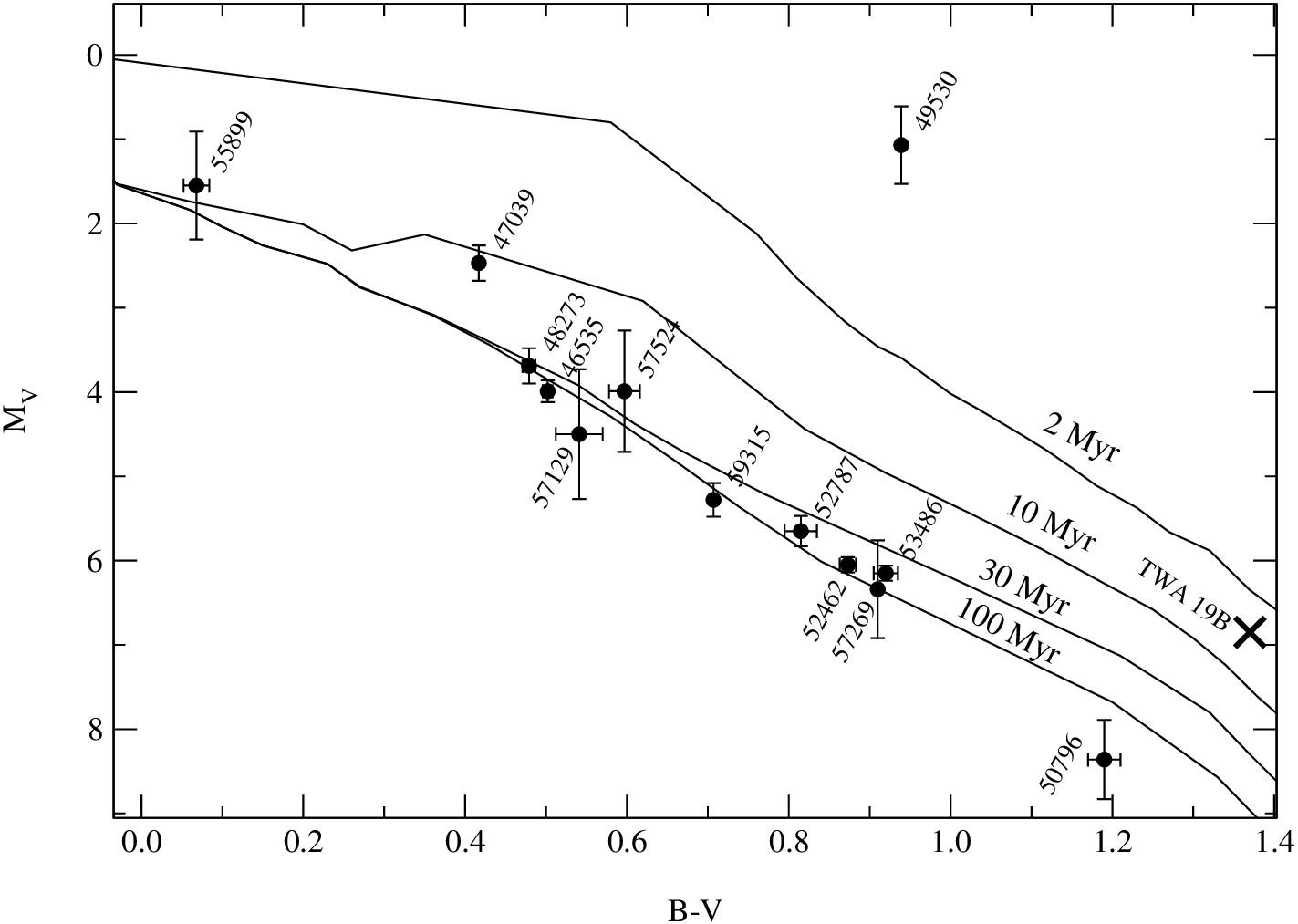}} \\
\end{tabular}\par}

\caption{Color magnitude diagram of Hipparcos MF candidates along with theoretical
isochrones from \citet{Siess} $\mathrm{Z}=0.02$
model. HIP~48273 and HIP~57269 are known binaries with almost equal
masses and the plotted points were corrected for this; that is, $M_v$
corresponds to that of one member. HIP 49530 is a giant
star (K1~III) and HIP 57129 is a W~UMa type contact binary \citep{Sistero}, thus
they cannot be members of TWA. TWA~19B (a K7 companion of HIP~57524) is  
plotted by using R \& I photometric data (R=11.06 and I=10.21) from \citet{TWA} 
and distance of the primary (104~pc).}
\end{figure}
We plotted absolute visual magnitude ($\mathrm{M}_{V}$) versus
$(B-V)$ values for all Hipparcos MF candidates where $(\mathrm{B}-\mathrm{V})$
values are either ground based measurements (taken from the Hipparcos catalog) or 
estimated from Tycho $(B_{T}-V_{T})$ values using a relation
given in Bessell (2000). The $\mathrm{M}_{V}$ values were calculated from 
ground-based or Tycho visual magnitudes and Hipparcos parallaxes.

From Figure~2, we can
eliminate most Hipparcos MF candidates except HIP~47039 and HIP~57524.
HIP~57524 is an already known TWA member \citep[TWA~19; ][]{TWA}. Although 
the location of HIP~57524 on a color magnitude diagram is not well defined due 
to a large error, its K7 companion (TWA~19B) confirms its very young age (Figure~2).
{\bfseries Based on a weak Li line and slow rotation, we believe 
HIP~47039 to be a post-ZAMS star; hence it cannot be a TWA member. }
HIP~48273 and HIP~57269 are binaries with about equal masses and
their absolute magnitudes were adjusted to account for this.
The age of HIP~55899 (A0V) cannot be estimated from Figure~2 because
an A0 star evolves very quickly so that in $\sim 10$~Myrs, A0 stars
arrive at the zero-age main-squence (ZAMS). Following \citet{AstarsZAMS},
we plotted all A-type stars with small errors in parallax and $(B-V)$
values from the Hipparcos catalog (Figure~3). Because of the fast evolution
of A-type stars, statistically, young A-type stars have to be located
at the bottom of the A-stars' distribution (ZAMS), as seen in Figure~3 for the
three young A-type stars (HR~4796A=TWA~11, HD~141569, and $\beta $~Pictoris).
HIP~55899 is not located on the ZAMS, thus it is unlikely to be as young 
as the other three stars. From Figures~1--3, we eliminated all candidates 
except HIP~57524 (TWA~19) and three Tycho stars (7760-0835-1, 8238-1462-1,
and 8234-2856-1).
\begin{figure}
{\centering \resizebox*{0.9\columnwidth}{!}{\includegraphics{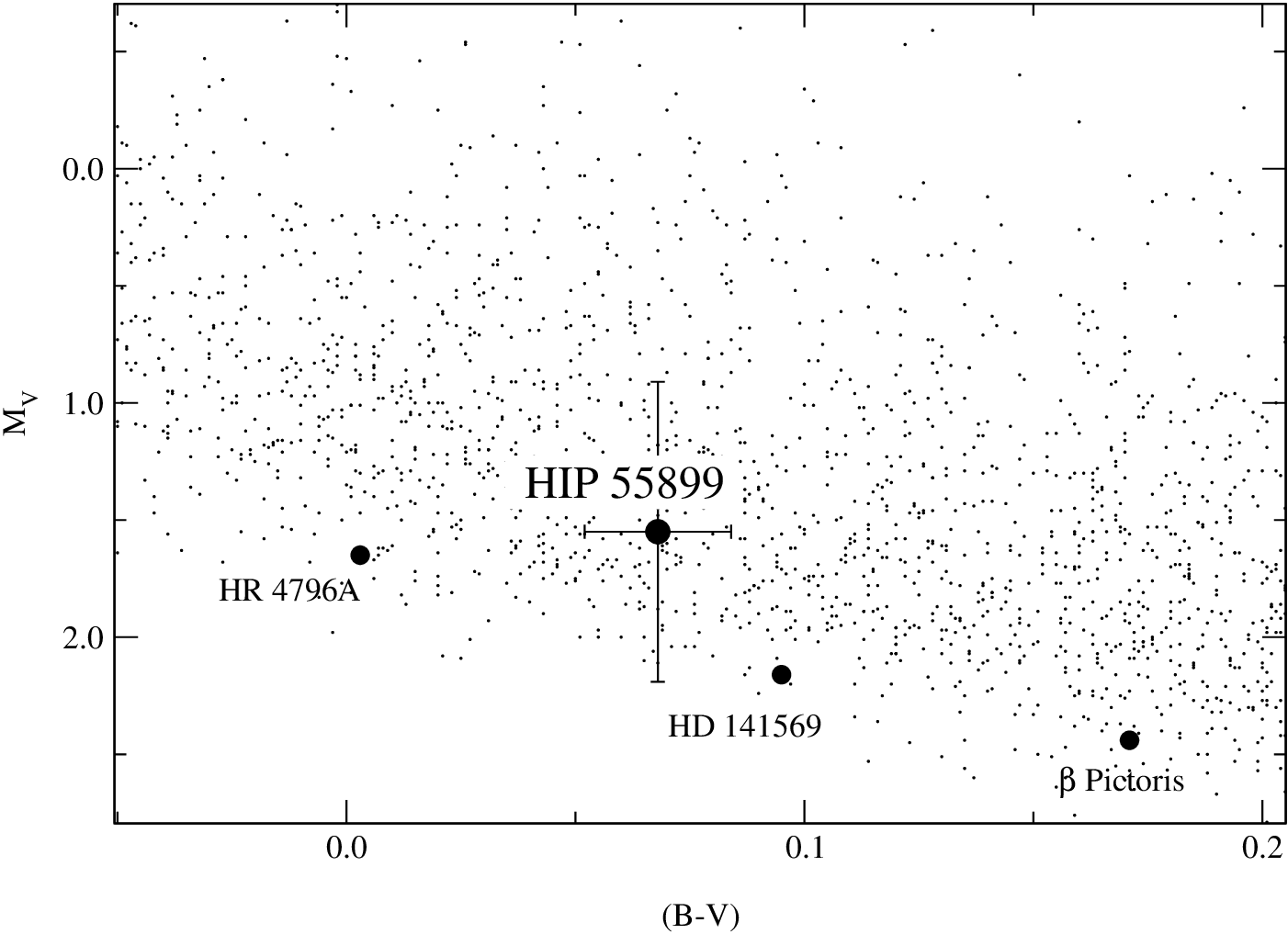}} \par}

\caption{A-type stars color magnitude diagram. Dots represent Hipparcos A-type
stars with small parallax error ($<10$~\%) and
small $(B-V)$ error ($<0.05$~mag).
HIP~55899 is plotted with error bars. Three well
known young ($\leq12$~Myrs) A-type stars are displayed as solid circles. }
\end{figure}

From their Li~6708~\AA{} strengths, positions in the sky (RA$\sim $12h,
DEC$\sim -45^{\circ }$), and proper motions ($\mu_\alpha\sim-34$~mas/yr and
$\mu_\delta\sim-12$~mas/yr), the three Tycho stars must be members of
the Lower Centaurus Crux (LCC) association. Assuming they are $\sim10$~Myr old stars, 
we can photometrically estimate their distance ($\sim130$~pc). 
All the known TWA members are likely $\leq$100~pc from Earth.
Now having eliminated all the distant TWA MF candidates (other than the 
three Tycho stars), we cannot confirm or 
reject a possible connection between TWA and the Sco-Cen complex.  Additional data,
especially identification of less massive members of the Sco-Cen complex,
are required to further investigate the possibility of such a connection.
Currently known members of the complex are mostly earlier than F/G-type
stars \citep{deZeeuw}. 
If TWA is not a part of the LCC or Sco-Cen
complex, based on the proximity of TWA~19 to LCC in space, it is
also possible that TWA~19 belongs to LCC rather than to TWA itself. 

In summary, among 23 \citet{MF} TWA candidates, we verify only three
possibly new TWA members (TYC~7760-0835-1, TYC~8238-1462-1, and
TYC~8234-2856-1). If TWA is not a part of a larger structure (the LCC or 
Sco-Cen complex), then these three Tycho stars may not be related to TWA. 
Therefore, MF's claims of an isotropic expansion of TWA and TWA being a 
part of the ``Gould-disk'' have to be re-evaluated based on our new results.
It is clearly shown that in identifying new members of nearby diffuse stellar 
associations or groups, one cannot rely solely on kinematic properties of 
candidates. When photometric 
-- putting the candidates in a color magnitude diagram -- and spectroscopic 
-- Li~6708\,\AA{} and H$\alpha$ -- data are combined
with kinematic information, much more convincing membership can be achieved 
as seen in \citet{bPic2}.

We thank Dr. van den Ancker for helpful remarks as the referee. This research
was supported in part by the UCLA Astrobiology Institute and by a NASA Origins
grant to UCLA.


\end{document}